\documentclass{article}
\usepackage{spconf,amsmath,graphicx}
\usepackage{bbding}
\usepackage{amssymb}
\usepackage{multirow}
\usepackage{support-caption}
\usepackage{subcaption}
\usepackage{indentfirst} 
\usepackage{url}

\newcommand{\rra}{RAR-U-Net}

\title{RAR-U-NET: a Residual Encoder to Attention Decoder by Residual Connections Framework for Spine Segmentation under Noisy Labels}
\name{Ziyang Wang$^1$,  Zhengdong Zhang$^2$, Irina Voiculescu$^1$}
\address{$^1$Department of Computer Science, University of Oxford, UK\\
$^2$ State Key Laboratory of Virtual Reality Technology and Systems, Beihang University, China}

\begin{document}
%\ninept
%
\maketitle
\begin{abstract}
Segmentation algorithms for medical images are widely studied for various clinical and research purposes. In this paper, we propose a new and efficient method for medical image segmentation under noisy labels. The method operates under a deep learning paradigm, incorporating four novel contributions. Firstly, a residual interconnection is explored in different scale encoders to transfer gradient information efficiently. Secondly, four copy-and-crop connections are replaced by residual-block-based concatenation to alleviate the disparity between encoders and decoders. Thirdly, convolutional attention modules for feature refinement are studied on all scale decoders. Finally, an adaptive denoising learning strategy (ADL) is introduced into the training process to avoid too much influence from the noisy labels. Experimental results are illustrated on a publicly available benchmark database of spine CTs. Our proposed method achieves competitive performance against other state-of-the-art methods over a variety of different evaluation measures.
% Computed Tomography scanners are commonly used for data acquisition, as the images are crucial for diagnosis, pre- and post- surgical assessment, and image-guided intervention. 
% The full implementation and trained models are publicly available.
%  This study deals with scan data environments which have been labelled noisily. That is, operators may have attempted to label parts of each scan, but there may be discrepancy between their opinion or their contour drawing precision.

\end{abstract}


\section{Introduction}
\label{sec:intro}
%Machine learning technologies, especially deep learning, are increasingly reliable for use applications such as medical image segmentation. 
The encoder-decoder model has been one of the most prominent deep neural network architectures used in medical image segmentation. U-Net~\cite{ronneberger2015u} is a completely symmetric variety of encoder-decoder. The encoder extracts pixel location features via down sampling; and the decoder recovers the spatial dimension and pixel location information with a deconvolution operation. Between the encoder and decoder layer, there is a copy-and-crop connection to deliver multi-scale information. U-Net has been used successfully for segmentation a variety of areas in the human body.
In 2016, Ronneberger proposed a 3D U-Net which carried out volumetric segmentation through extracting sparsely annotated volumetric images~\cite{cciccek20163d}. This is achieved at the cost of an increase in the number of training parameters. In 2018 Oktay~\cite{oktay2018attention} proposed an attention gate model to enable convolutional neural networks automatically to learn target structures of different shapes and sizes. It is based on Attention U-Net which achieves higher sensitivity and accuracy while requiring minimal computational overhead. In 2019, Residual Networks, Inception Networks, Densely Connected Networks and several modified U-Net have continued to be explored and optimised~\cite{guan2019fully,diakogiannis2020resunet,zhou2018unet++}. The high resolution Dense-U-Net Network for spine segmentation proposed by Kola{\v{r}}{\'\i}k et al.~\cite{kolavrik2019optimized} explores the performance of 2D and 3D U-Nets in residual networks and densely connected networks. Due to the 3D convolutional layers and interconnections, this architecture is costly in training parameter time.

Meanwhile, a separate hurdle when segmenting medical data is the potential lack of precision in the annotated contours, usually due to limitations in knowledge and to clinicians' subjectivity. Results of the annotation process can depart from the the gold standard: labelled features can present slight erosion or dilation of what would be the ideal contours, as well as various kinds of elastic transformations. We hereafter call such variations `noisy labels', which will affect the effect of the final model.

In this work, we overcome the shortcomings described above through  \rra, a {\bf R}esidual encoder to {\bf A}ttention decoder by {\bf R}esidual connections framework for medical image segmentation under noisy labels. Its main novelty consists of:

1) shortcut interconnections on the four down-sampling blocks as residual encoders, to enhance gradient information transfer,

2) residual-block-based concatenation to mitigate the disparity between encoders and decoders,

3) convolutional attention module on four up-sampling blocks to capture essential information, 

4) adaptive denoising learning strategy which eliminates the negative effects of noise labels in the training process.

% The adaptive learning approach allows clean noisy label while training process. \rra\ is tested, compared with other classical models, with ablation studies, and evaluated with a large collection of metrics. The results achieve competitive performance compared with other state-of-the-art approaches.

\section{Methods}

\begin{figure*}  
\centering  
\includegraphics[width=0.85\linewidth]{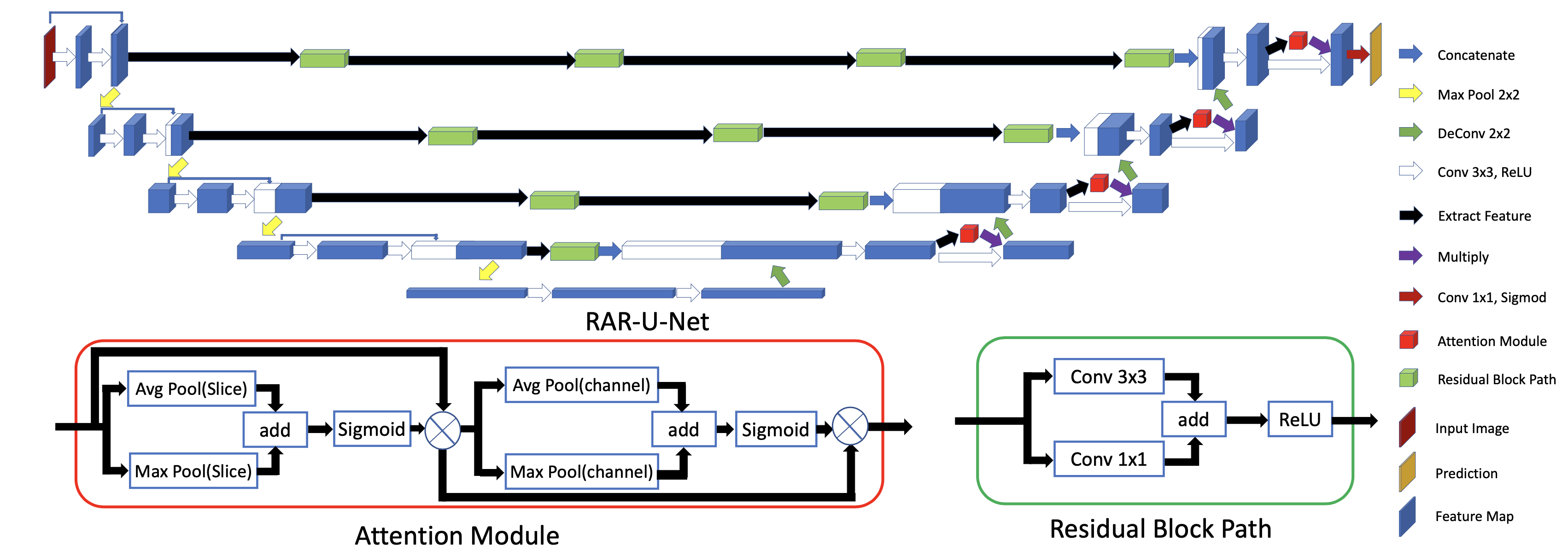}  
\caption{The Architecture of the Proposed Network, Attention Module and Residual Block Path}  
\label{fig:picture001}  
\end{figure*}

The architecture of \rra\ is illustrated in Fig.~\ref{fig:picture001}. It is a symmetrical architecture which consists of convolution, upsampling and downsampling, allowing to contract and recover pixel-level information. 
We detail below the four specific contributions of this method.

\subsection{Residual Interconnections Networks} \label{2.1}
Inspired by ResNet~\cite{he2016deep}, a concatenation function is allocated to each down-sampling block. In order to strengthen the ability to express features and gradient information, after obtaining the downsampled features of a previous layer, these features will go through a new feature extraction process. The extraction process consists of the repeated application of two $3 \times 3$ unpadded convolutions, each followed by a rectified linear unit (ReLU). The number of feature channels is doubled compared to the previous layer after the first $3 \times 3$ convolution operator. Finally, we concatenate the downsampled feature with the feature acquired from the second $3 \times 3$ convolution operator. This operation aims to establish connections between different layers, making full use of feature information and alleviating the gradient disappearance problem.

\subsection{Residual-Block-Based Concatenation}
\label{2.2}

To mitigate the disparity between encoders and decoders which may degrade the segmentation performance~\cite{ibtehaz2020multiresunet}, the four copy and crop connections are replaced by residual-block-based concatenation. We adopt residual learning to connect the encoder to the decoder in each layer. 

%A building block is shown in Fig.~\ref{fig:picture001} which is named as Residual Block Path. Formally, in this paper we define a building block in Equation~\ref{residual}.
We define the Residual Block Path as a building block of our model, also sketched in Fig.~\ref{fig:picture001}. Formally, if $x$ and $y$ are the input and output vectors of the layers, then
\begin{equation}\label{residual}
y = F(x,\left\{{W_i}\right\})+x
\end{equation}
where the Eq.~\ref{residual} illustrates residual block, and $F(x,\left\{{W_i}\right\})$  represents the residual mapping to be learned. For example in Fig.~\ref{fig:picture001} that has one $3 \times 3$ layer, $F = \sigma {W_1}x$ in which $\sigma$ and $W$ denote the ReLU activation and weight, to simplify notation, the biases are omitted. The operation $F + x$ is performed by a shortcut connection and element-wise addition. We also use a $1 \times 1$ convolution operator to match the number of channels. 
After the addition operation we adopt the second non-linearity ReLU. Finally, the feature map processed through the residual block is concatenated with the upsampled feature of the decoder. To mitigate the difference between each level of the encoder and decoder, and considering the computational cost, the number of Residual Blocks at each level of the encoder-decoder is set to 4, 3, 2 and 1, respectively.

\subsection{Convolutional Attention Module}
\label{2.3}

% . It aims to increase weight of feature map on important features and suppressing unnecessary area and noisy label. 

To enhance the performance of the decoder classification for each pixel by capturing essential information in the presence of noisy labels, we explore the use of an attention mechanism. Unlike attention gate filter features from skip connections~\cite{oktay2018attention}, an attention module can normally be integrated with convolutional layers to enhance key information of the feature map with pooling layers and sigmoid activation functions~\cite{woo2018cbam}. 

Our proposed attention module for the convolutional layer of decoder is sketched in Fig.~\ref{fig:picture001}. There are two parts of the attention module related to the channel-- and spatial attention of different feature maps. Both parts are developed by the pooling layer and sigmoid activation. Average and max pooling layers avoid noisy label gradients to keep trunk parameters. The sigmoid function simultaneously generates a weight attention value as output for each pixel location and channel. 

Fig.~\ref{fig:picture001} shows how a feature map $F \in R^{W\times H\times C}$ with the size of shape $W\times H\times C$ from a previous CNN is sent to the attention module pipeline. 
The feature maps from the average pooling layers and max pooling layers of spatial dimensions $W\times H$  are denoted $F^{Avg}_{Spatial}$ and $F^{Max}_{Spatial} \in R^{W\times H\times 1}$. Similarly, $F^{Avg}_{Channel}$ and $F^{Max}_{Channel} \in R^{1\times 1\times C}$  are the feature maps from average pooling layers and max pooling layers on the channel dimension $C$. 

Both of the spatial attention value ${W_{Spa}}$ and channel attention value ${W_{Channel}}$ are calculated through the sigmoid activation $\sigma$. The final output feature map $F_{out}$ is adaptively refined from feature map $F$ through a spatial attention layer and channel attention layer successively, in order to capture essential information.

%The final output of the weight attention value $W$ is calculated from the above feature maps through the sigmoid activation $\sigma$. It can then be multiplied with specific features to capture essential information.

\begin{eqnarray} \small
\label{attention}
W_{Spa}   =   \sigma  (F^{Avg}_{Spa} + F^{Max}_{Spa})
\end{eqnarray}
\begin{eqnarray} \small
W_{Channel}   =   \sigma (F^{Avg}_{Channel} +  F^{Max}_{Channel})
\end{eqnarray}
\begin{eqnarray} \small
F_{out}   =   W_{Channel}(W_{Spa}(F) \otimes F) \otimes (W_{Spa}(F) \otimes F)
\end{eqnarray}
% \begin{eqnarray} \small
% W   =   \sigma  (F^{Avg}_{Spa} + F^{Max}_{Spa}) + 
% \sigma (F^{Avg}_{Channel} +  F^{Max}_{Channel}) 
% \end{eqnarray}

\begin{figure}[ht]
\centering
\begin{subfigure}{.15\textwidth}
  \centering
  % include first image
  \includegraphics[width=\linewidth]{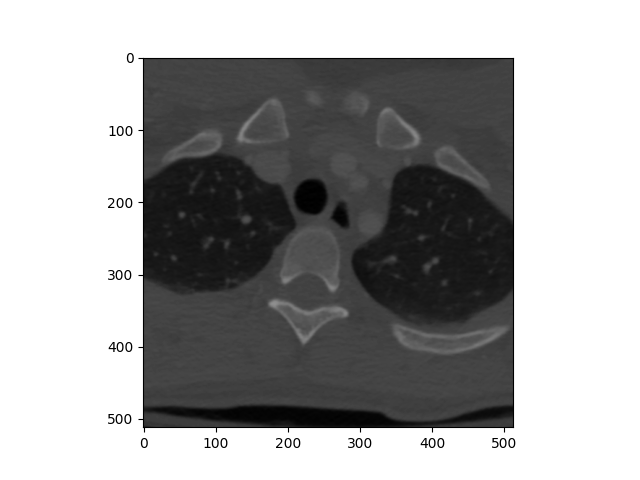}  
  \caption{Raw image}
  \label{fig:a}
\end{subfigure}
\begin{subfigure}{.15\textwidth}
  \centering
  % include second image
  \includegraphics[width= \linewidth]{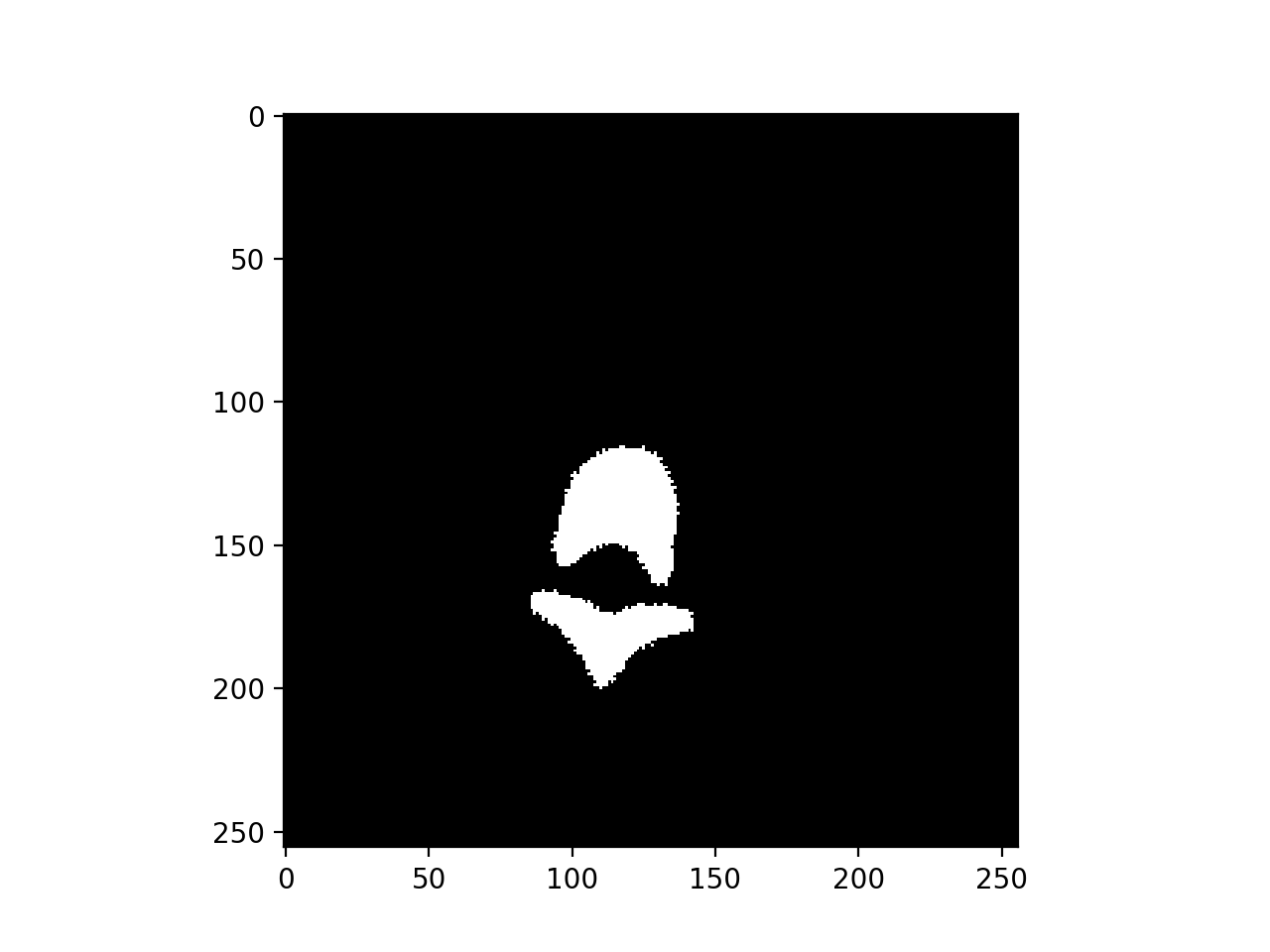}  
  \caption{GT mask}
  \label{fig:b}
\end{subfigure}
\begin{subfigure}{.15\textwidth}
  \centering
  % include first image
  \includegraphics[width= \linewidth]{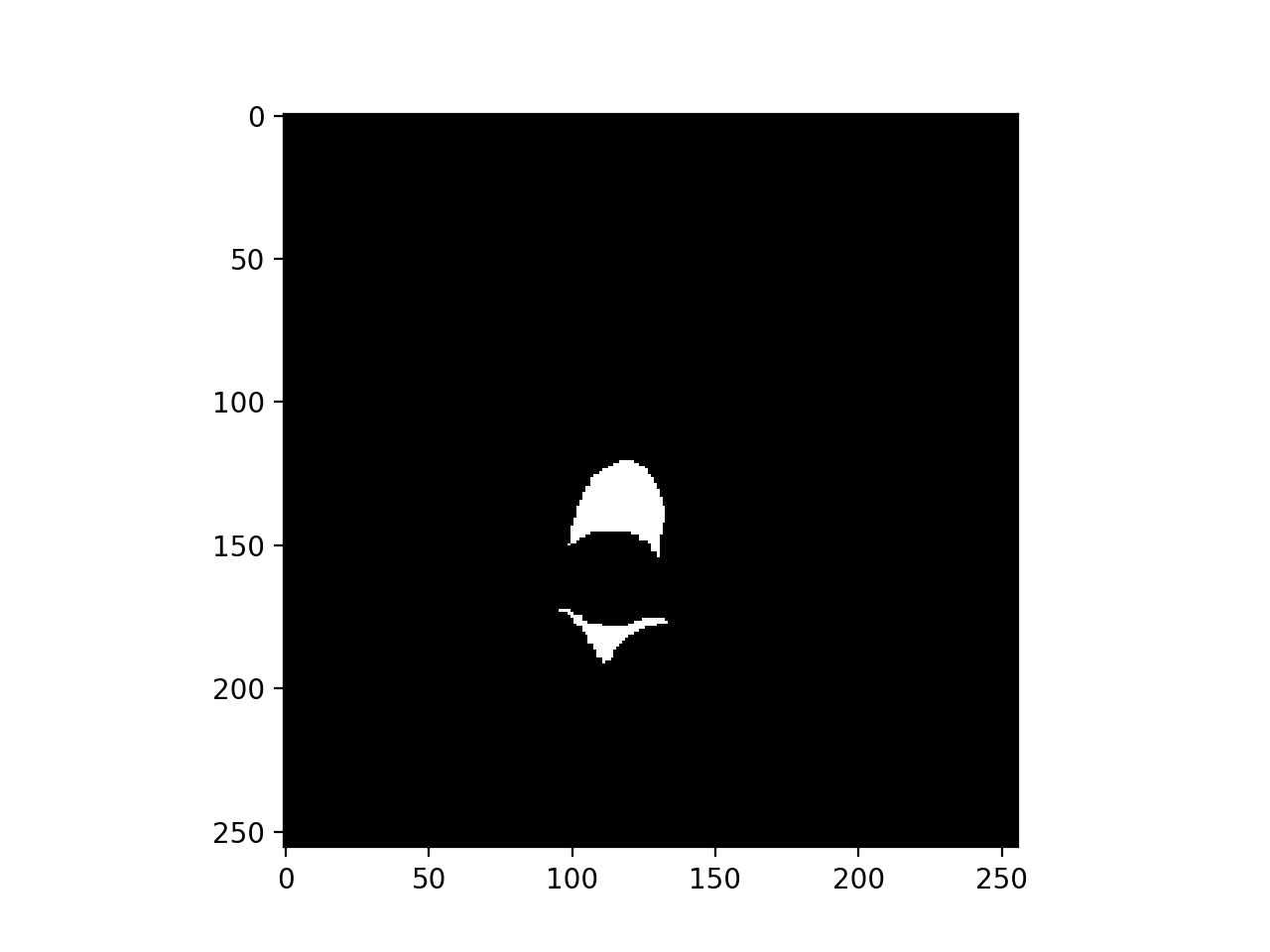}  
  \caption{Erosion}
  \label{fig:c}
\end{subfigure}
\begin{subfigure}{.15\textwidth}
  \centering
  % include second image
  \includegraphics[width= \linewidth]{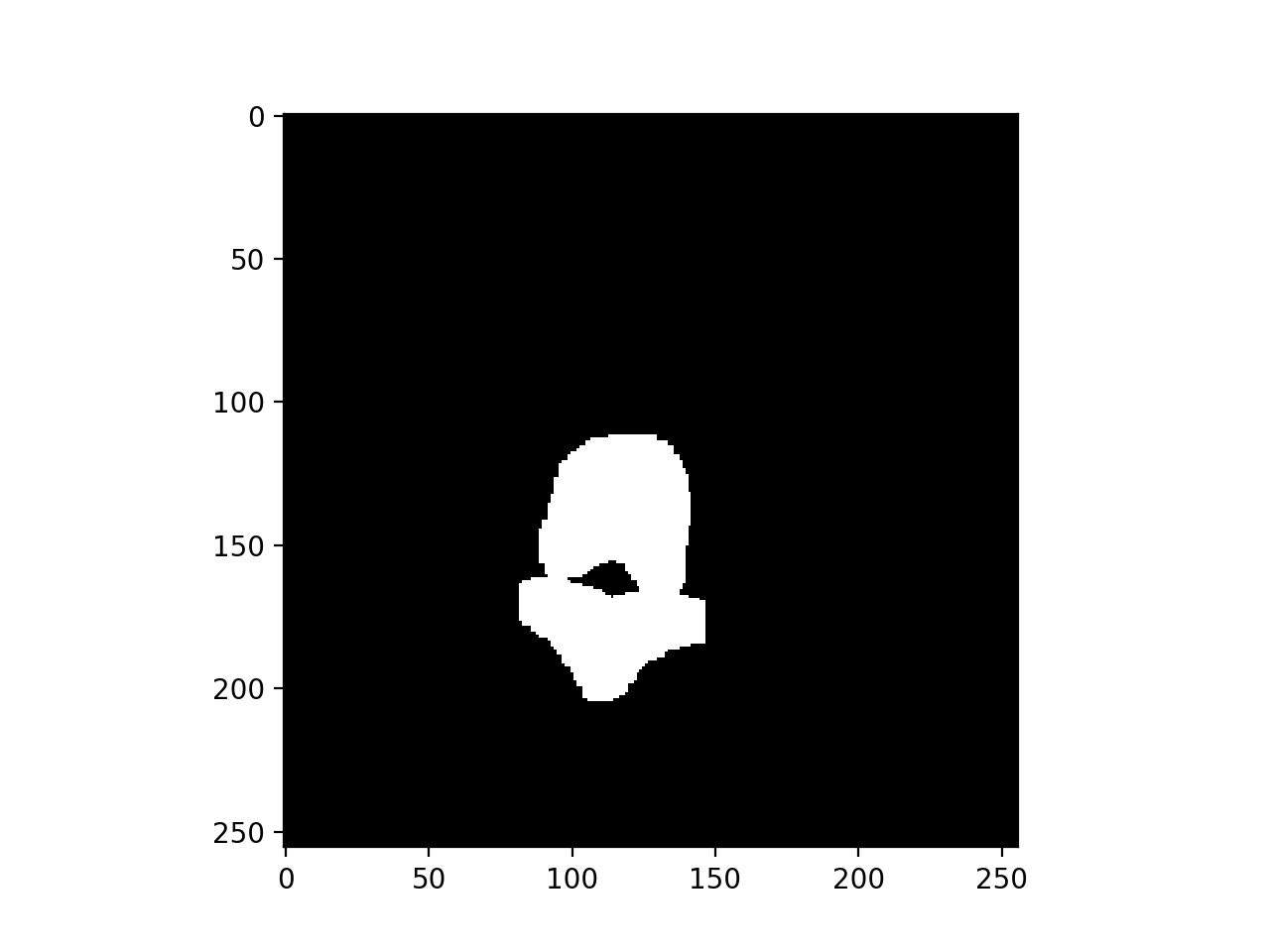}  
  \caption{Dilation}
  \label{fig:d}
\end{subfigure}
\begin{subfigure}{.15\textwidth}
  \centering
  % include second image
  \includegraphics[width= \linewidth]{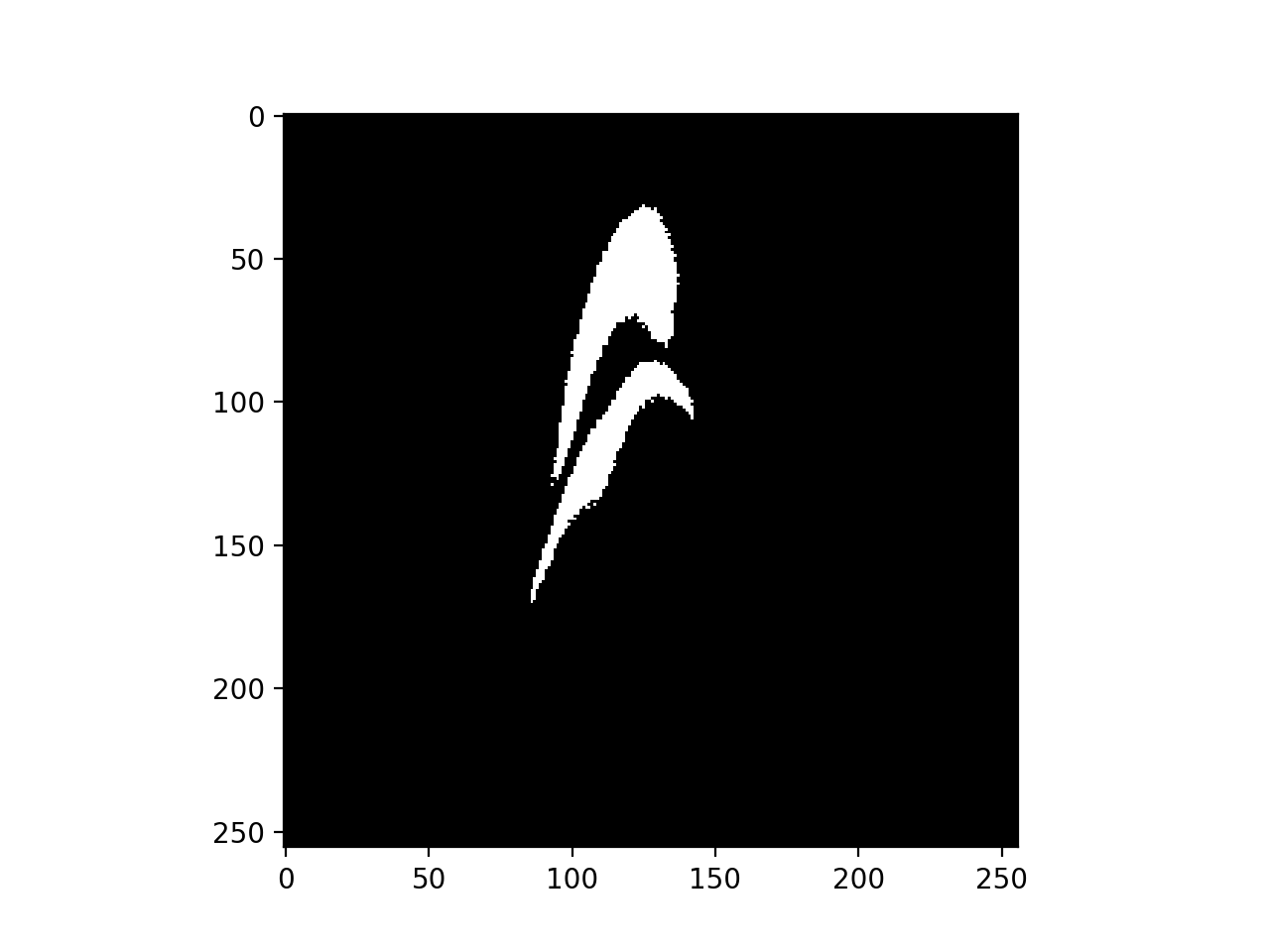}  
  \caption{Elastic transform}
  \label{fig:e}
\end{subfigure}
\begin{subfigure}{.15\textwidth}
  \centering
  % include second image
  \includegraphics[width= \linewidth]{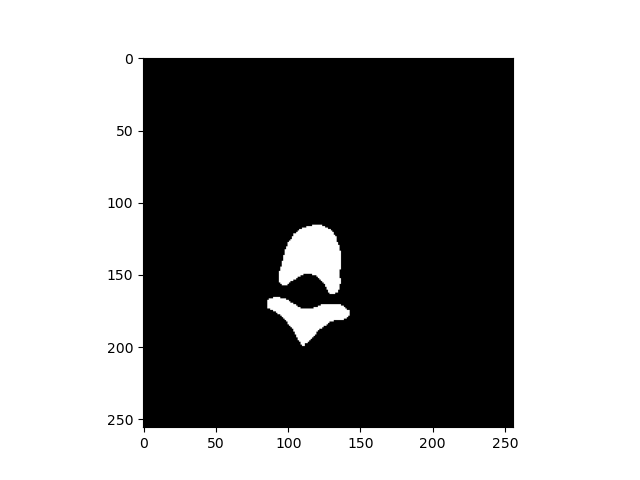}  
  \caption{Predicted result}
  \label{fig:f}
\end{subfigure}
\caption{Examples of Spine CT Slice and its Alterations}
\label{fig:fig}
\end{figure}

\subsection{Adaptive Denoising Learning}
\label{2.4}

%Deep learning training in clinical environment are suffering from the problem of inaccuracy labeled data. Different operators can label a same CT scan with different masks or even result in wrong label with noisy label. 
In order to simulate the challenge of noisy (or inaccurate) labels emerging from practical settings, we propose a strategy of adaptive denoising to be applied during the training process. 
%Masks in the training dataset are randomly selected and manually replaced as noisy labels. 
To simulate this situation, a certain proportion $\beta$ of masks in the training data have been replaced with synthetically generated noisy labels. These labels present with erosions, dilations or elastic transforms. The extent of the noise present is denoted by $\alpha \in [0,1]$.
Three such examples are illustrated in Fig.~\ref{fig:c}--\ref{fig:e}, which are noisy in some sense compared to the ground truth in Fig.~\ref{fig:b}. 
%Two parameters for noisy dataset including: a) the noisy level of noisy label $\alpha \in (0,1)$, b) the proportion of noisy label $\beta$ are both defined and test in this experiment. 
The noise level $\alpha$ can be thought of as the extent of the overlap between the original ground truth mask and the generated noisy label. 

We propose a simple and efficient adaptive denoising learning strategy. 
%This is different from detecting and cleaning noise before training.  
Inspired by O2U-Net~\cite{huang2019o2u}, the difference between each prediction and label data is calculated and recorded at every training epoch. The higher the loss of a label, the higher its probability of being a noisy label. %The proposed learning strategy allows to detect and remove a specific number of label with high loss value during iteration. 
During training, our strategy aims to detect and remove a number of high loss value labels.
A large number of noisy labels get detected at the begining of the training iteration, and then a few more towards the end.
This is because the training process evolves from underfitting to overfitting. The number $N(t)$ of labels detected and removed in each epoch is
\begin{equation}\label{cleannoisylabel}
N(t) = \left\{
            \begin{array}{lr}
             0.5 (1-\alpha) \beta y , & 0 < t <0.1 (1-\alpha) \beta x \\
             \frac{-y}{x}t + 0.6 (1-\alpha) \beta y, & \hspace{-3mm}0.1 (1-\alpha) \beta x \leq t \\
             & \& ~~ t \leq 0.5 (1-\alpha) \beta x\\
             0.1 (1-\alpha) \beta y , &  0.5 (1-\alpha) \beta x < t \leq x
             \end{array}
\right.
\end{equation}
% ORIGINAL EQUATION BELOW
\iffalse
\begin{equation}
\small
N(t) = \left\{
            \begin{array}{lr}
             0.5 (1-\alpha) \beta y , & 0 < t <0.1 (1-\alpha) \beta x \\
             \frac{-y}{x}t + 0.6 (1-\alpha) \beta y, & \hspace{-3mm}0.1 (1-\alpha) \beta x \leq t \leq 0.5 (1-\alpha) \beta x\\
             0.1 (1-\alpha) \beta y , &  0.5 (1-\alpha) \beta x < t \leq x
             \end{array}
\right.
\end{equation}
\fi
where $t$ is the current training epoch, $\alpha$ is the noise level, $\beta$ is the proportion of items in the training dataset to which noise has been applied, $x$ is the total number of training epochs, and $y$ is the total number of masks. 0.1 and 0.5 are hyperparameters obtained using a systematic search.
%, after many comparative experiments.

\section{Experiments and Results}
\setlength{\tabcolsep}{2pt}
\begin{table}[t] \small
\begin{center}
%\caption{Segmentation Results on Spine CT Dataset}
\caption{Direct Comparison Against Existing Algorithms}
\label{table:overlap}

\begin{tabular}{llllllr}
\hline\noalign{\smallskip}
Model & Dice & Acc & Pre & Rec & Spe & \hspace{-5pt}Par \tiny{$10^{6}$} \\
\noalign{\smallskip}
\hline
\noalign{\smallskip}
UNet  & 0.8360  & 0.9863 & 0.8832 & 0.7936 & 0.9952 & 7.26\\
Residual-UNet &  0.8810  & 0.9898 & 0.9097 & 0.8540 & 0.9961 & 9.90\\
Densely-UNet & 0.8316  & 0.9860 & 0.8832 & 0.7857 & 0.9952 & 15.47\\
M-UNet & 0.9478 & 0.9954 & 0.9512 & 0.9444	& 0.9978 & 7.77\\
M-Densely-UNet & 0.9517 & 0.9958 & 0.9524 & 0.9508 & 0.9978 & 15.48\\
VGG16 UNet & 0.9138 & 0.9925 & 0.9235 & 0.9043 & 0.9966 & 23.75\\
VGG19 UNet & 0.9024 & 0.9914 & 0.9029 & 0.9019 & 0.9955 & 29.06\\
ResNet34 UNet & 0.6626 & 0.9689 & 0.6333 & 0.6947 & 0.9815 & 24.45\\
SE-ResNet34 UNet & 0.7306 & 0.9762 & 0.7265 & 0.7347 & 0.9873 & 24.61\\
ResNeXt101 UNet & 0.7597 & 0.9765 & 0.6909 & 0.8438 & 0.9826 & 32.06\\
DenseNet121 UNet & 0.7982 & 0.9811 & 0.7526 & 0.8498 & 0.9872 & 12.13\\
InceptionV3 UNet & 0.8109 & 0.9837 & 0.8250 & 0.7972 & 0.9922 & 29.93\\
% MobilenetV2 UNet & 0.5671 & 0.9586 & 0.5240 & 0.6179 & 0.9742 \\
EfficientNet UNet & 0.8358 & 0.9857 & 0.8431 & 0.8286 & 0.9929 & 10.11\\
MultiRes UNet & 0.8542 & 0.9864 & 0.8094 & 0.9043 & 0.9902 & 7.76\\
3D UNet & 0.8078 & 0.9874 & 0.7788 & 0.8390 & 0.9922 & 22.58\\
3D Residual-UNet & 0.7757 & 0.9850 & 0.7360 & 0.8198 & 0.9904 & 28.15 \\
3D Densely-UNet & 0.7921 & 0.9860 & 0.7450 & 0.8456 & 0.9906 & 44.78\\
3D Attention UNet & 0.8623 & 0.9870 & 0.8129 & 0.9182 & 0.9902 & 22.60\\
LinkNet & 0.8958 & 0.9908 & 0.8919 & 0.8999 & 0.9950 & 20.32 \\
FPN & 0.8804 & 0.9893 & 0.8675 & 0.8936 & 0.9937 & 17.59\\
{\bf \rra} & {\bf 0.9580 } & {\bf  0.9963} & {\bf 0.9605 }& {\bf 0.9554 }& {\bf 0.9982} & 11.79\\
\hline
\end{tabular}
\end{center}
\end{table}

\subsection{Dataset and Experimental Setup}

We used a publicly available spine dataset from California and NIH~\cite{yao2012detection}, which consists of CT scans from 10 patients, of up to 600 slices per scan, at a resolution of $512\times512$, and $1$mm inter-slice spacing. All images are normalized and resized to $256 \times 256$. Ground truth (GT) masks are available for each image, some of which were used as input for the noise introduction. 
Data augmentation was applied in the form of $\pm 90^\circ$ rotations. 
Of the 10 scans, 9 were used for training and 1 for testing. Validation is carried out on 10\% of the training data. An overlapping approach allowing to add eight more surrounding slices in each training batch is utilized to ensure the 3D model collect continuous information.

% [Overlapping is more relevant to training a 3D. Consider leaving out this detail if not including 3D algorithms. If including 3D, then explain more about how this was done.]

\begin{table}[t]\small
\setlength{\abovecaptionskip}{0.1cm}
\setlength{\belowcaptionskip}{-0.1cm}
\caption{Difference Measures of the Segmentation Results}
\label{tab:dif}
\begin{center}
\centering
\begin{tabular}{ |c|c|c|c| }
\hline
 &  HD (pixels) & ASSD (pixels) & RVD $\in[0,1]$ \\
 \hline 
 \rra & 10.0110 & 0.8221 & 0.0518 \\
  \hline
\end{tabular}
\end{center}
\end{table}

\begin{table}[t]\small
\setlength{\abovecaptionskip}{0.1cm}
\setlength{\belowcaptionskip}{-0.1cm}
\caption{Boundary-based Match in the Segmentation Results}
% TODO: only retain the first line of this table
\label{tab:sbd}
\begin{center}
\centering
\begin{tabular}{ |c|c|c|c| }
\hline
 &  DBD$_G$  & DBD$_M$  & SBD  \\
 \hline 
 \rra & 0.8425 & 0.8564 & 0.8465 \\
%Dice & 0.8425 & 0.8564 & 0.8465 \\
% Jaccard & 0.7712 & 0.7930 & 0.7792 \\
%TPVF & 0.8274 & 0.8636 & 0.8420 \\
%TNVF & 0.7824 & 0.6754 & 0.7298 \\
%FPVF & 0.2176 & 0.3245 & 0.2702 \\
%FNVF & 0.1726 & 0.1364 & 0.1579 \\
% Precision & 0.8989 & 0.8764 & 0.8853 \\
  \hline
\end{tabular}
\end{center}
\end{table}

\setlength{\tabcolsep}{2pt}
\begin{table}[t]\small

\caption{Ablation Studies on Contributions of Architecture}
\label{tab:ablationcomparison}
\begin{center}
\centering
\begin{tabular}{ |c|c|c|c|c|c|c| }
\hline
\multirow{2}{4em}{Residual Encoders} & \multirow{2}{5em}{Residual Connections} & \multirow{2}{4em}{Attention Decoders} & \multirow{2}{3em}{IOU} &  \multirow{2}{3em}{Recall} & \multirow{2}{5em}{Trainable Parameters}\\
 & &  &  & &\\
 \hline
  & & & 0.7182 & 0.8832 &  7,762,465\\
\checkmark & & & 0.7873 & 0.8540 &  9,899,625\\
 & \checkmark & & 0.9119 & 0.9549 & 8,912,673\\
 &  & \checkmark &  0.8927 & 0.9406 & 7,785,157\\
\checkmark &  & \checkmark &  0.9070 & 0.9481 & 9,922,317\\
\checkmark & \checkmark &  & 0.9126 & 0.9535 & 11,049,833 \\
\checkmark & \checkmark & \checkmark & {\bf 0.9193} & {\bf 0.9605} & 11,794,125\\ 
 \hline
\end{tabular}
\end{center}
\end{table}
\setlength{\tabcolsep}{2pt}

\begin{table}[t]\small

\caption{Ablation Studies on ADL}
\label{tab:ablationnoisy}
\begin{center}
\centering
\begin{tabular}{ |c|c|c|c|c|c| }
\hline
Proportion & Level & Algorithm & ADL & IOU & Recall \\
 \hline
75\% & 0.68 & U-Net & & 0.6445 & 0.7303 \\
75\% & 0.68 & U-Net & \checkmark & {\bf 0.6742}& {\bf 0.8072} \\
 \hline
75\% & 0.68 & Residual-UNet & & 0.7732 & 0.9097 \\
75\% & 0.68 & Residual-UNet & \checkmark & {\bf 0.8138} & {\bf 0.9462} \\
 \hline
75\% & 0.68 & Attention-UNet & & 0.7809 & 0.8823\\
75\% & 0.68 & Attention-UNet & \checkmark & {\bf 0.8087} & {\bf 0.9142}\\
  \hline
50\% & 0.77 & UNet & & 0.7523 & 0.8420\\
50\% & 0.77 & UNet & \checkmark & {\bf 0.8522} & {\bf 0.9295} \\
  \hline
50\% & 0.77 & Attention-UNet & & 0.8464 & 0.9201\\
50\% & 0.77 & Attention-UNet & \checkmark & {\bf 0.8561} & {\bf 0.9283} \\
  \hline
25\% & 0.85 & Residual-UNet & & 0.8615 & 0.9051\\
25\% & 0.85 & Residual-UNett & \checkmark & {\bf 0.8868} & {\bf 0.9433}\\
  \hline
25\% & 0.85 & Dense-UNet & & 0.8443 & 0.9378\\
25\% & 0.85 & Dense-UNett & \checkmark & {\bf 0.8864} & {\bf 0.9424}\\
  \hline
25\% & 0.55 & U-Net & & 0.8024 & 0.8698\\
25\% & 0.55 & U-Net & \checkmark & {\bf 0.8304} & {\bf 0.9176}\\
  \hline
25\% & 0.55 & Residual-UNet & & 0.8230 & 0.8956\\
25\% & 0.55 & Residual-UNet & \checkmark & {\bf 0.8495}  & {\bf 0.9126} \\
  \hline
\end{tabular}
\end{center}
\end{table}

%\subsection{Experimental Setup}

The \rra\ code was developed in Python using Tensorflow~\cite{tensorflow2015-whitepaper}. It has been run
% under Ubuntu 18.04.1
on an Nvidia GeForce RTX2080 Ti GPU with 16GB memory, and Intel(R) Xeon(R) CPU E5-2650 v4. The runtimes varied between 1000--1200 mins for 50 epochs, including the data transfer. With a training batch size of 8, the learning rate is $10^{-5}$.
% Rather than accuracy, the Dice coefficient is selected as loss function, because of the imbalance between background and spine.
%Given the imbalance between the spine and background pixels, t
The loss function is based on the Dice coefficient. 
%The total training epochs is 50, a choice described in Section~\ref{2.4}. %Some of the benchmarks are developed by an open source library~\cite{Yakubovskiy:2019}.
% , which has been made available publicly~\cite{zywcode}.

\subsection{Results and Discussion}

% [Discuss that the spine forms a small part of the overall image. This means a perverse algorithm could label everything black, and in this case Accuracy would still be potentially high. The aim, therefore, is to minimize the number of false positives. This is why Dice was used in training. And this is why Dice or Recall are more relevant to the evaluation process than just Accuracy. Talk about Recall and Precision as a pair.]

% [From Google: Choose the right loss function Binary crossentropy might lead your network in the direction of optimizing for all labels, now if you have an unbalanced amount of labels in your image, it might draw your network to just give back either white, gray or black image predictions. Try using the dice coefficient loss]

Figs.~\ref{fig:a},~\ref{fig:b}, and~\ref{fig:f} illustrate examples of a raw image, GT mask and average predicted result. \rra\ is compared with classical segmentation algorithms including LinkNet~\cite{chaurasia2017linknet}, FPN~\cite{kim2018parallel}, 3D UNet~\cite{cciccek20163d}, MultiResUnet~\cite{ibtehaz2020multiresunet}, DenselyUnet~\cite{kolavrik2019optimized}, and U-Net with classical backbones such as VGG~\cite{simonyan2014very}, ResNet\cite{he2016deep}, SE-ResNet~\cite{hu2018squeeze}, ResNeXt~\cite{xie2017aggregated}, InceptionV3~\cite{szegedy2016rethinking} and EfficientNet~\cite{tan2019efficientnet}.
% To evaluate the performance, we have assessed our predicted masks against the GT masks available with the dataset.
We first compare the performance of our algorithm against a collection of widely used overlap measures such as the Dice coefficient, Accuracy, Precision, Sensitivity (or Recall), Specificity, which enable the comparison against other methods. These are illustrated in Table~\ref{table:overlap}, together with the number of training parameters. The proportion of noisy labels in this case is 0.

The performance of our algorithm through difference measures such as the Hausdorff Distance, Average Symmetric Surface Distance (ASSD) and Relative Volume Difference (RVD) is shown in Table~\ref{tab:dif} -- the smaller, the better.

Table~\ref{tab:sbd} reports the extent to which the boundaries of the machine segmentation MS match those of the GT~\cite{yeghiazaryan2018family}, using the Directed Boundary Dice relative to GT (DBD$_G$), Directed Boundary Dice relative to MS (DBD$_M$) and Symmetric Boundary Dice (SBD). In a von Neumann neighbourhood $N_x$ of each pixel $x$ on the boundary $\partial G$ of the ground truth,
\begin{equation}
\small
DBD_G=DBD(G, M)=\displaystyle\frac
    {\rule{0pt}{3ex}
    \sum\limits_{x \in \partial G} \text{Dice}(N_x)}
    {\left| \partial G \right|}
\end{equation}
\begin{equation}
\small
    SBD =
    \displaystyle
    \frac
    {\rule{0pt}{3ex}
    \sum\limits_{x \in \partial G} DSC(N_x) + \sum\limits_{y \in \partial M} DSC(N_y)}
    {\left| \partial G \right| + \left| \partial M \right|}
\end{equation}
%The symmetric average is being brought down by DBD$_G$ when the latter features isolated areas of false negative labels (also identified by the FNVF). 
These measures penalise mislabelled areas in the machine segmentation. Even a 75\% close match between the boundaries is considered a good result. 
%Seen that the boundary match is an asymmetric (directional) measure, DBD$_M$ is invariably better. 
%Table~\ref{tab:sbd} reports these.

\subsection{Ablation study}

In order to analyze the effects of each of the four proposed contributions and their combinations, extensive ablation experiments have been conducted. Table~\ref{tab:ablationcomparison} documents how the removal of one or more components compromises the overall performance. The same table also gives a measure of the complexity of the overall \rra\ model and its sub models. 
Table~\ref{tab:ablationnoisy} focuses specifically on the effect of the ADL strategy. `Proportion' and `Level' illustrate how many images are randomly chosen to be processed as a noisy label and their level of noise in the training dataset. These have been varied more than shown here. `Algorithm' illustrates that different algorithms are separately utilized for training; the adaptive denoising level `ADL' strategy enables all models to perform better under noisy labels.  

% such as: 2D-U-Net~\cite{ronneberger2015u}, 2D Densely-U-Net~\cite{kolavrik2019optimized}, MultiRes-U-Net~\cite{ibtehaz2020multiresunet}, LinkNet~\cite{chaurasia2017linknet}, FPN~\cite{lin2017feature} and several original U-Net with backbones such as VGG16~\cite{simonyan2014very}, InceptionV3~\cite{szegedy2016rethinking}. 
% , yet the symmetric average is being brought down by DBD$_G$ when the latter features isolated areas of false negative labels (also identified by the FNVF). 

% Dice, accuracy, precision, recall and specificity are illustrated in Equation~\ref{dice}, \ref{Accuracy}, \ref{Precision}, \ref{Sensitivity}, \ref{Specificity} where TP, TN, FP, FN means True Positive, True Negative, False Positive, and False Negative~. A large of boundary overlap evaluation methods are also used~\cite{yeghiazaryan2018family}. 

% NEXT BIT COMMENTED OUT
% \iffalse
% \begin{equation}
% Dice = \frac{2TP}{2TP+FP+FN}\label{dice}
% \end{equation}
% \begin{equation}
% Accuracy = \frac{TP+TN}{TP+TN+FP+FN}\label{Accuracy}
% \end{equation}
% \begin{equation}
% Precision = \frac{TP}{TP+FP}\label{Precision}
% \end{equation}
% \begin{equation}
% Recall = \frac{TP}{TP+FN}\label{Sensitivity}
% \end{equation}
% \begin{equation}
% Specificity = \frac{TN}{TN+FP}\label{Specificity}
% \end{equation}
% \fi

\section{Conclusions}

%In this study, a novel framework for medical image segmentation is proposed and studied. E
Our experimental results demonstrate that all four proposed contributions significantly improve segmentation under noisy labels, at a smaller training parameter cost. Although the tests were specific to a single public dataset, the methods are generic.
%In this study, a framework for medical image segmentation is proposed and studied. Comprehensive evaluations and comparisons are completed, and \rra\ achieves promising performance. 
%In the future, semi-supervised learning based on \rra\ with unlabeled data will be studied.
% Residual shortcut connection is explored in encoder to improve performance with low computational cost. Connection between encoder and decoder is also explored based on residual block. Attention mechanism is also designed and utilized in decoder to improve performance. Adaptive learning is designed which is easy to implement and efficient. 

\bibliographystyle{IEEEbib}  
\bibliography{refsabridged}

\begin{thebibliography}{10}

\bibitem{ronneberger2015u}
O~Ronneberger et~al.,
\newblock ``{U-Net}: Convolutional networks for biomedical image
  segmentation,''
\newblock in {\em Int Conf Med Im Comp \& Comp-Assisted Intervention}.
  Springer, 2015, pp. 234--241.

\bibitem{cciccek20163d}
{\"O}zg{\"u}n {\c{C}}i{\c{c}}ek, Ahmed Abdulkadir, Soeren~S Lienkamp, Thomas
  Brox, and Olaf Ronneberger,
\newblock ``{3D U-Net}: learning dense volumetric segmentation from sparse
  annotation,''
\newblock in {\em Int Conf Med Im Comp \& Comp-Assisted Intervention}.
  Springer, 2016, pp. 424--432.

\bibitem{oktay2018attention}
O~Oktay et~al.,
\newblock ``Attention {U-Net}: Learning where to look for the pancreas,''
\newblock {\em Int Conf Medical Imaging with Deep Learning}, 2018.

\bibitem{guan2019fully}
Steven Guan, Amir~A Khan, Siddhartha Sikdar, and Parag~V Chitnis,
\newblock ``Fully dense unet for 2-d sparse photoacoustic tomography artifact
  removal,''
\newblock {\em IEEE journal of biomedical and health informatics}, vol. 24, no.
  2, pp. 568--576, 2019.

\bibitem{diakogiannis2020resunet}
Foivos~I Diakogiannis, Fran{\c{c}}ois Waldner, Peter Caccetta, and Chen Wu,
\newblock ``Resunet-a: a deep learning framework for semantic segmentation of
  remotely sensed data,''
\newblock {\em ISPRS Journal of Photogrammetry and Remote Sensing}, vol. 162,
  pp. 94--114, 2020.

\bibitem{zhou2018unet++}
Zongwei Zhou, Md~Mahfuzur~Rahman Siddiquee, Nima Tajbakhsh, and Jianming Liang,
\newblock ``Unet++: A nested u-net architecture for medical image
  segmentation,''
\newblock in {\em Deep Learning in Medical Image Analysis and Multimodal
  Learning for Clinical Decision Support}, pp. 3--11. Springer, 2018.

\bibitem{kolavrik2019optimized}
M~Kola{\v{r}}{\'\i}k et~al,
\newblock ``Optimized high resolution 3d dense-u-net network for brain and
  spine segmentation,''
\newblock {\em Applied Sciences}, vol. 9, no. 3, pp. 404, 2019.

\bibitem{he2016deep}
K~He et~al,
\newblock ``Deep residual learning for image recognition,''
\newblock in {\em Proc IEEE CVPR}, 2016, pp. 770--778.

\bibitem{ibtehaz2020multiresunet}
Nabil Ibtehaz and M~Sohel Rahman,
\newblock ``Multiresunet: Rethinking the u-net architecture for multimodal
  biomedical image segmentation,''
\newblock {\em Neural Networks}, vol. 121, pp. 74--87, 2020.

\bibitem{woo2018cbam}
S~Woo et~al.,
\newblock ``{CBAM}: Convolutional block attention module,''
\newblock in {\em Proc ECCV}, 2018, pp. 3--19.

\bibitem{huang2019o2u}
J~Huang et~al.,
\newblock ``{O2U-Net}: A simple noisy label detection approach for deep neural
  networks,''
\newblock in {\em Proc IEEE ICCV}, 2019, pp. 3326--3334.

\bibitem{yao2012detection}
J~Yao et~al,
\newblock ``Detection of vertebral body fractures based on cortical shell
  unwrapping,''
\newblock in {\em Int Conf Med Im Comp \& Comp-Assisted Intervention}.
  Springer, 2012, pp. 509--516.

\bibitem{tensorflow2015-whitepaper}
Mart\'{\i}n Abadi and etc,
\newblock ``{TensorFlow}: Large-scale machine learning on heterogeneous
  systems,'' 2015,
\newblock Software available from tensorflow.org.

\bibitem{chaurasia2017linknet}
Abhishek Chaurasia and Eugenio Culurciello,
\newblock ``Linknet: Exploiting encoder representations for efficient semantic
  segmentation,''
\newblock in {\em 2017 IEEE Visual Communications and Image Processing (VCIP)}.
  IEEE, 2017, pp. 1--4.

\bibitem{kim2018parallel}
Seung-Wook Kim, Hyong-Keun Kook, Jee-Young Sun, Mun-Cheon Kang, and Sung-Jea
  Ko,
\newblock ``Parallel feature pyramid network for object detection,''
\newblock in {\em Proceedings of the European Conference on Computer Vision
  (ECCV)}, 2018, pp. 234--250.

\bibitem{simonyan2014very}
Karen Simonyan and Andrew Zisserman,
\newblock ``Very deep convolutional networks for large-scale image
  recognition,''
\newblock {\em Int Conf Learning Representations(ICLR)}, 2015.

\bibitem{hu2018squeeze}
Jie Hu, Li~Shen, and Gang Sun,
\newblock ``Squeeze-and-excitation networks,''
\newblock in {\em Proceedings of the IEEE conference on computer vision and
  pattern recognition}, 2018, pp. 7132--7141.

\bibitem{xie2017aggregated}
Saining Xie, Ross Girshick, Piotr Doll{\'a}r, Zhuowen Tu, and Kaiming He,
\newblock ``Aggregated residual transformations for deep neural networks,''
\newblock in {\em Proceedings of the IEEE conference on computer vision and
  pattern recognition}, 2017, pp. 1492--1500.

\bibitem{szegedy2016rethinking}
Christian Szegedy, Vincent Vanhoucke, Sergey Ioffe, Jon Shlens, and Zbigniew
  Wojna,
\newblock ``Rethinking the inception architecture for computer vision,''
\newblock in {\em Proc IEEE CVPR}, 2016, pp. 2818--2826.

\bibitem{tan2019efficientnet}
Mingxing Tan and Quoc Le,
\newblock ``Efficientnet: Rethinking model scaling for convolutional neural
  networks,''
\newblock in {\em International Conference on Machine Learning}. PMLR, 2019,
  pp. 6105--6114.

\bibitem{yeghiazaryan2018family}
V~Yeghiazaryan et~al.,
\newblock ``Family of boundary overlap metrics for the evaluation of medical
  image segmentation,''
\newblock {\em SPIE JMI}, vol. 5, no. 1, pp. 015006, 2018.

\end{thebibliography}

\end{document}